\documentstyle[prd,aps,floats,epsfig]{revtex}

\newcommand{\lav}{\langle}
\newcommand{\rav}{\rangle}
\newcommand{\be}{\begin{equation}}
\newcommand{\ee}{\end{equation}}
\newcommand{\beq}{\begin{equation}}
\newcommand{\eeq}{\end{equation}}
\newcommand{\ba}{\begin{eqnarray}}
\newcommand{\ea}{\end{eqnarray}}
\newcommand{\dle}[1]{\label{#1}}
\newcommand{\dla}[1]{\label{#1}}
\newcommand{\dr}[1]{\ref{#1}}
\newcommand{\dc}[1]{\cite{#1}}

\newcommand{\hti}{\tilde{h}}

\newcommand{\gsim}{\raise.3ex\hbox{$>$\kern-.75em\lower1ex\hbox{$\sim$}}}
\newcommand{\lsim}{\raise.3ex\hbox{$<$\kern-.75em\lower1ex\hbox{$\sim$}}}

\newcommand{\k}{\ell}

\newcommand{\gzt}{g(z,t)}

\newcommand{\gzpt}{g(z',t)}

\newcommand{\Ai}{A(z;z',z-z')}
\newcommand{\Aiii}{A(z+z';z,z')}
\newcommand{\Bii}{B(z,z'-z;z')}
\newcommand{\Biiii}{B(z',z-z';z)}

\newcommand{\zetac}{\zeta_c}
\newcommand{\tchi}{\tilde{\chi}}

\newcommand{\GGmu}{\Gamma G \mu}

\title{The statistical physics of cosmological networks of string loops}

\begin{document}
\author{Jo\~ao Magueijo$^a$, H\aa vard Sandvik$^a$, and
Dani\`ele A.\ Steer$^b$}
\address{{\it a})
Blackett Lab., Imperial College, Prince Consort Road, London 
SW7 2BZ, U.K.\\ 
{\it b}) D.A.M.T.P., Silver Street, Cambridge, CB3 9EW, U.K.} 

\twocolumn[\hsize\textwidth\columnwidth\hsize\csname@twocolumnfalse\endcsname 
 
\maketitle

\begin{abstract}
We solve numerically the Boltzmann equation describing the evolution
of a cosmic string network which contains only loops.  
In Minkowski space time the equilibrium solution predicted by
statistical mechanics is recovered, and we prove that this solution is
stable to non-linear perturbations provided that their energy 
does not exceed the critical energy for the Hagedorn transition. 
In expanding Einstein - de Sitter Universes we probe the
distribution of loops with length much smaller than the horizon.  For
these loops we discover stable scaling solutions both in the radiation
and matter dominated epochs.  The shape of these solutions is very 
different in the two eras, with
much higher energy density in the radiation epoch, and a larger
average loop length in the matter epoch. These results suggest that
if the conditions for formation of loop networks are indeed satisfied,
these could in principle be good candidates for structure formation. 
\end{abstract}

]

\typeout{--- Title page start ---}

\renewcommand{\thefootnote}{\fnsymbol{footnote}}


\section{Introduction}

One of the greatest challenges of modern cosmology is to explain
the large scale structure of the universe:  how did the universe, 
which is known to have been very homogeneous at early times,
eventually develop structures such as galaxies and anisotropies in the cosmic
microwave background (CMB)?
There are currently two different classes of theory which try to
provide an explanation---inflation \cite{infl} and topological
defects \cite{csreviews}---though the two may also be combined 
\cite{CHM2,stinf1,stinf2}.
In inflationary scenarios, initial vacuum quantum fluctuations
eventually lead to the formation of structures. In 
topological defect theories a network of vacuum defects is 
produced at a phase transition in the early universe.
These defects naturally provide a  source
of inhomogeneity for all subsequent times.

Although there are several classes of defects \cite{Neil,Ruth}, 
currently the most popular candidate for structure formation is
cosmic strings. These
are line-like vacuum defects fully specified
by a single parameter, $\mu$, their energy per unit length
\cite{James1,CHM,avelino,James2,natalie,tanmay}.   
For strings to be viable candidates for structure formation, they
should neither come to dominate the energy density of the universe nor 
disappear altogether. Instead they should ``scale'', by which we
mean that the energy density in strings should remain 
a fixed fraction of the energy density in the universe.

The formation and evolution of a network of cosmic strings is
extremely complex.   A large amount of 
analytical~\cite{ACK,ACK2,EdRome,CJAP,dp}
and numerical~\cite{AT,BB,AS,Mark} 
work has been devoted to exploring the 
evolution of these networks in expanding flat universes with no
cosmological constant.  
Current wisdom is that the network quickly
evolves towards a fixed scaling solution, 
though the details of this solution
differ between the radiation and matter eras.
In this attractor
solution the network is dominated by infinite strings.

In a widely known but unpublished manuscript, Ferreira and Turok
\dc{Pedro} were the first
to conjecture that a different scaling solution might be
reached if only string loops were formed during the phase transition. 
In this different scaling solution no infinite strings are present,
given that none exist in the initial conditions: the network
is forever made up solely of loops.
This  conjecture is an intriguing possibility, with potential
far reaching consequences for structure formation.  Undoubtedly 
the time-time correlators of
the stress energy tensor of such networks is rather different.
Radically different predictions for galaxy clustering and 
CMB anisotropies could ensue.

It was later argued that networks containing only loops might indeed 
form after first order string producing phase transitions 
\cite{borrill,bubble,bubble1}. Such transitions proceed via bubble
nucleation, and as the bubbles expand and collide, string loops are produced.
It was suggested that this mechanism may be vastly 
more effective than the more conventional Kibble mechanism. 

In this paper we use the model of Ref.\ ~\cite{dp}
to study the evolution of networks of string loops through a
Boltzmann-type  rate equation.  Previously this equation was studied
analytically in detail in references \dc{dp,dthesis}.  However, given the
complicated nature of the equation --- it is a non-linear and non-local,
integro-partial differential equation --- only asymptotic solutions
were searched for,  that is, the focus was on the behaviour of loops
whose length was much bigger than the horizon size.  It was argued
\dc{dp} that a stable scaling solution  exists in Minkowski
space time as long as the initial energy density in loops is less than 
a critical value which corresponds to the Hagedorn energy.
In a radiation dominated universe, a scaling solution was also found.
However, this was thought to be unstable --- depending on the initial
energy density in loops, these were either thought to disappear, or
infinite strings were predicted to form.  In the matter era, no stable 
solution was found, and for much of parameter space the loops were
expected to disappear.

The purpose of this paper is to search for numerical 
solutions of the full rate equation of \dc{dp}, rather 
than asymptotic solutions.  This is a
challenging numerical task given the complicated nature of the rate
equation, and in fact we are only able to probe accurately the
behaviour of small loops.  The reasons for this and some of our
methods for tackling the numerical
problems are described in the rest of this paper.  We also
describe the results of our code and  elaborate on the 
properties of the solutions found.

This paper is organised in the following way.  In Section~\ref{rate} we
review the rate equation and set up our notation.
In Section~\ref{numerical}
we discuss some further analytical massaging connected with 
the practical aspects of our numerical implementation.
The core of our paper is in Sections~\ref{mink} and \ref{expand},
where we present our numerical solutions, as well as some analytical
results used as a check upon the numerics.  The results presented
there can be summarised as follows. 
In a Minkowski background we recover the scaling solution predicted by
statistical mechanics. We prove that this solution is stable
against non linear perturbations, as long as these do not exceed
the critical mass for the Hagedorn transition. 
In expanding Universes and for loops much smaller than the horizon, 
we discover stable 
scaling solution both in the radiation and matter dominated
epochs.
The shape of the solution is quite different in the two eras, with
much higher energy density in the radiation epoch, and a larger
average loop length in the matter epoch.  
Based on these results, we suggest that
if the conditions for formation of loop networks are indeed satisfied, 
these could in principle be good candidates for structure formation.
Finally, in Section~\ref{conc} we conclude, outlining further
work prompted by our results.

\section{The Rate Equation}\label{rate}

In this section we set up our notation and review the rate equation 
proposed in  \cite{dp} for the evolution of networks of loops.  For
clarity we introduce the non-expanding and expanding universe cases
separately.  The rest of the paper will focus on searching numerically for 
solutions of the equation.

\subsection{Non-expanding Universe}

Consider first the evolution of a network of cosmic string loops in a
non-expanding universe.  In this case one would expect the network to 
reach an equilibrium, time-independent distribution. 
We will study the approach to equilibrium and the propties of
that solution.

Let ${n(\ell,t)}$ be the number of loops per unit volume with physical
length between $\k$ and $\k+d\k$ at a time $t$. In \cite{dp} it was
argued that in a non-expanding Universe $n(\ell,t)$ satisfies 
\ba
\frac{\partial n}{\partial t}
& = & 
 \int_{0}^{\ell/2}A(\ell;\ell',\ell-\ell')n(\ell',t)n(\ell-\ell',t)d\ell'
\nonumber\\
& + &  \int_{\ell}^{\infty} B(\ell,\ell'-\ell;\ell')n(\ell',t)d\ell'
\nonumber \\
& - & {n(\ell,t)}   \int_{0}^{\infty} A(\ell+\ell';\ell,\ell')n(\ell',t)d\ell'
\nonumber \\
& - & {n(\ell,t)}   \int_{0}^{\ell/2}  B(\ell',\ell-\ell';\ell)d\ell'
,
\dla{ratemink}
\ea
where the four integrals describe collision interactions between the
loops.  The coefficient $A(\ell+\ell';\ell,\ell')$
gives the probability that two separate loops of length $\ell$ and
$\ell'$ collide and produce a loop of length $\ell+\ell'$ (assuming
as usual that in the collision partners are exchanged),
and it is given by \dc{dp}
\begin{equation} 
A(\ell_1+\ell_2;\ell_1,\ell_2) = \chi \ell_1 \ell_2
\label{Acoef},
\end{equation}
where $\chi$ is a relative velocity.
The function $B(\ell',\ell-\ell';\ell)$ gives the probability that
a loop of length $ \ell$ self-intersects to produce two loops of length 
$\ell'$ and $\ell-\ell'$.  Arguments in \dc{dp,dthesis} suggested that
\begin{equation}
B(y,\ell-y;\ell) = \frac{\tilde{\chi} y (\ell-y) (\ell+\zeta)^{5/2}}
{\bar{\xi}^{3/2}
(\ell-y+\zeta)^{5/2}(y+\zeta)^{5/2}},
\label{Bcoef}
\end{equation}
where $\tilde{\chi}$ is another relative velocity.  The two lengths in 
(\dr{Bcoef}) are $\bar{\xi}$, the distance along which the Brownian
loops are correlated in direction, and
$\zeta$, the distance between any kinks on the loops.  
These lengths are assumed to be time
independent in a non-expanding universe. 

The total length density of loops, $L$ is given by
\begin{equation}
	L(t) = \int_{0}^{\infty}n(\k,t) \k \; d\k
\dle{Ldef}
\end{equation}
and in Minkowski spacetime this remains constant since energy is
conserved.  In fact the proof of this, substiting the rate equation 
into the time derivative of (\dr{Ldef}) is quite complex as one often
has to deal with diverging integrals \dc{dthesis}.  This problem of
infinities will also have to be dealt with in our numerical work,
where we shall use (\dr{Ldef}) as a check on our
code. 

The total loop number density is given by
\begin{equation}
	N(t) = \int_{0}^{\infty}n(\k,t) \; d\k.
\end{equation}
There are no limitations on this quantity, but combined with the
energy $E \sim L$ it gives information on the shape of the
loop distribution.

\subsection{Rate equation in Expanding-Universe}

In an expanding Universe the rate equation becomes \dc{dp}:
\begin{eqnarray}
\frac{\partial n}{\partial t}& = & 
-{n(\ell,t)}  \frac{\partial {\dot{\ell}}}{\partial \ell} -
\frac{\partial {n(\ell,t)} }{\partial \ell} {\dot{\ell}} + \lim_{\ell \rightarrow
0^+}\left[n(\ell,t) \dot{\ell}
\right] \delta(\ell) \nonumber\\&-& 3H{n(\ell,t)} \nonumber
 \\
 & +&
 \int_{0}^{\ell/2}A(\ell;\ell',\ell-\ell')n(\ell',t)n(\ell-\ell',t)d\ell'
\nonumber
 \\
& + &  \int_{\ell}^{\infty} B(\ell,\ell'-\ell;\ell')n(\ell',t)d\ell'
\nonumber\\
& - & {n(\ell,t)}   \int_{0}^{\infty} A(\ell+\ell';\ell,\ell')n(\ell',t)d
\ell'
\nonumber\\
& - & {n(\ell,t)}   \int_{0}^{\ell/2}  B(\ell',\ell-\ell';\ell)d\ell',
\dla{rateeqn}
\end{eqnarray}
where $\dot{\ell}$ is the rate of change of the length of a loop.  This
has contributions from expansion, redshift and gravitational
radiation, and may be written as \dc{dp}
\begin{equation}
\dot{\ell} = - \Gamma G \mu + \frac{K\ell}{t},
\label{ldotfull}
\end{equation}
where we expect $0< K \lsim 0.1$ \dc{dp,dthesis}.  The third term in
(\dr{rateeqn}) guarantees that $n(\ell,t)=0$, $\forall \ell < 0$
$\forall t$, whilst the fourth term is the contribution from the
expansion of the universe with $H$ being the Hubble parameter.
The functions $A$ and $B$ in (\dr{rateeqn}) 
are the same as those of equations
(\dr{Acoef}) and (\dr{Bcoef}) respectively, where
we again assume the same two length scales on the loops, $\bar{\xi}$
(distance along which loops are correlated in direction---c.f.\
3-scale model \dc{ACK}) and
$\zeta$ (the distance between kinks---c.f.\ \dc{ACK}).  Now though
these two scales are assumed to `scale' \dc{dp}:
\begin{equation}
\bar{\xi} = \bar{\xi}_c t \; \; \; \; \; \zeta = \zetac t
\dle{scaling}
\end{equation}
where $\bar{\xi}_c$ and $\zetac$ are both constants.
As discussed in \dc{dp} and in analogy with \dc{ACK} we expect 
\begin{equation}
\zetac \simeq \GGmu \bar{\xi}_c .
\dle{zetasize}
\end{equation}

Our aim will be to see whether the assumption (\dr{scaling}) combined with
equation (\dr{rateeqn}) are consistent with the energy
density in loops also scaling.
For that reason it is convenient to rewrite the rate equation
(\dr{rateeqn}) in the scaling variables
\ba
z & = & \frac{\ell}{t}
\label{zdef}
\\
 \gzt & = & {t^4} {n(\ell,t)},
\label{gdef}
\ea
so that scaling energy density corresponds to $g(z,t)=g(z)$.
Recall that from the definitions of 
$A$ (\dr{Acoef}) and $B$ (\dr{Bcoef}) one has
\begin{eqnarray}
A(zt;z't,(z-z')t) &= &t^2 \Ai,
\\
B(zt;(z'-z)t,z't) & = & \frac{1}{t^2} \Bii
\end{eqnarray}
where it is important to remember that in
$\Bii$, $\bar{\xi}$ must be replaced by $\bar{\xi}_c$. 
Hence the rate equation for $\gzt$ becomes:
\begin{eqnarray}
t \frac{\partial \gzt}{\partial t}& = & 
\gzt \left( 4 - \frac{3}{p} - K \right) \nonumber \\
&+&
\frac{\partial \gzt}{\partial z} \left( z(1-K) + \GGmu \right)
\label{ga}\nonumber 
\\
& - &
\GGmu \delta(z) [\gzt]|_{z \rightarrow 0^+}\nonumber 
\nonumber \\
 & +
& \int_{0}^{z/2} \Ai \gzpt g(z-z',t)dz'
\nonumber  \\
& + &  \int_{z}^{\infty} \Bii \gzpt dz'
\nonumber  \\
& - & \gzt  \int_{0}^{\infty} \Aiii \gzpt dz'
\nonumber  \\
& - & \gzt  \int_{0}^{z /2}  \Biiii dz'
\label{geqn}.
\end{eqnarray}
Here the parameter $p$ is defined by
the expansion rate; $R \propto t^{1/p}$, so that $p=2$ in the
radiation  era, and $p=\frac{3}{2}$ in the matter era. 
Observe that the RHS of the above equation is not explicitly $t$
dependent and hence that it is consistent to have
$\gzt = g(z)$ -- {\em i.e.}\ a scaling solution is consistent
with the equation.

Our aim is to solve equation (\dr{geqn}) to check whether such a
scaling solution is indeed reached, and if so what are its properties.

\section{Numerical implementation}\label{numerical}

\subsection{Variables suitable for numerical work}
\label{sec:newvar}

Any numerical approach to solve equation (\dr{geqn}) has to 
first deal with the obvious numerical problem posed by 
the semi-infinite interval $[0,\infty)$ in which the $z$
variable lies.  As was discussed in \dc{dp,dthesis}, just chopping off 
the integrals in (\dr{geqn}) at some upper value of $z$ is {\em not} a 
resolution to this problem
 as this changes completely the nature of the solutions of the 
rate equation.  Instead we introduce 
a new variable:
\begin{equation}
	y(z) = \frac{z}{1+z/c_{max}}
\label{ydef}
\end{equation}
in which $c_{max}$ is a fixed number which will be chosen to optimise
our numerical analysis.  Note that, as opposed to $z$, the
variable $y$ now lies in the finite range $0 \leq y \leq c_{max}$.
Also it will be useful to write down the inverse of (\dr{ydef});
\be
z(y) = \frac{y}{1-y/c_{max}}
\dle{zofy}
\ee

We then define 
\begin{equation}
	\hti(y(z),t)=g(z,t)
\end{equation}
so that the boundary condition $g(z \rightarrow \infty,t)=0$ $\forall
t$ becomes
\begin{equation}
	\hti(c_{max},t)=0 \; \; \forall t,
\end{equation}
and may now be imposed without loss of generality.   This
boundary condition is necessary so that the energy density in
loops is finite.

Furthermore, in an {\em expanding universe only}, 
we change variables to logarithmic time
\beq
	t \rightarrow u = \ln(t),
\eeq
so that on letting $h(y,u)=\hti(y,t(u))$ the rate equation
becomes
\begin{eqnarray}
&&\frac{\partial h(y,u)}{\partial u}= h(y,u) \left( 4 - \frac{3}{p} -
K \right) \nonumber \\
&& +  \frac{\partial h(y,u)}{\partial y} (1-y/c_{max})^2\left(
z(1-K) + \GGmu \right)
\nonumber
\\
&& - 
\GGmu \delta(y) [h(y,u)]|_{y \rightarrow 0^+}
\nonumber
\\
 && +
 \int_{0}^{\frac{y}{2-\frac{y}{c_{m}}}} A(z(y'),z(y)-z(y')) h(y',u)
\nonumber
 \\
&& \; \; \; \; \; \; \times
h(y(z(y)-z(y')),u)\frac{ dy'}{(1-y'/c_{max})^2} 
\nonumber
 \\
&& +   \int_{y}^{c_{max}} h(y',u)\left[ B(z(y),z(y'))- h(y,u) \right.
\nonumber
 \\
&& \; \; \; \; \; \; \times
\left. A(z(y),z(y')) \right]
\frac{dy'}{(1-y'/c_{max})^2}
\nonumber
 \\
&& -  h(y,u)  \int_{0}^{y} A(z(y),z(y'))h(y',u)
\frac{dy'}{(1-y'/c_{max})^2} 
\nonumber
 \\
&& -  h(y,u) \int_{0}^{\frac{y}{2-y/c_{max}}} B(z(y'),z(y))
\frac{dy'}{(1-y'/c_{max})^2} 
\label{h4}.
\end{eqnarray}
Note that for clarity we have now omitted the unnecessary arguments of
the $A$ and $B$ coefficients:  for example, we have written
$A(z;z',z-z')=A(z',z-z')$ and $B(z,z'-z;z')=B(z,z'-z)$.


In a non-expanding universe the rate equation is that given in
(\dr{h4}) but where one retains only the four integral terms on the
RHS, and where there is no logarithmic time (so $t=u$).

\subsection{Choice of parameters}

In order to solve (\dr{h4}) numerically, we clearly have to choose
values for the various parameters which appear in the rate equation as 
well as a value for $c_{max}$.

Firstly, by comparison with numerical simulations of cosmic string
evolution in expanding universes (see the discussion in section 3 of
\dc{dp} and also \dc{ACK}), we have taken
\be
	\chi = \tchi = 0.2
\dle{chival}
\ee
though in general these relative velocities need not be equal.  The
values given in (\dr{chival}) were also used for our simulations of
loops in Minkowski space.

The correlation length $\bar{\xi}_c$ of equation (\dr{scaling}) was
taken to be
\beq
	\bar{\xi}_c = 1
\eeq
which is of the order of the horizon size.  Note that the only place 
in which $\bar{\xi}_c$ appears in the rate 
equation is in the coefficient $B(z',z;z+z')$, where the relevant
combination on constants is $\tilde{\chi}/\bar{\xi}_c^{3/2}$. 
Thus we can change the
size of $\bar{\xi}_c$ simply by scaling $\tilde{\chi}$.
The value of $\zeta_c$ is approximately given by (\dr{zetasize}), so
we have taken
\beq	
	\zeta_c =5.0 \times 10^{-5}
\eeq
since for GUT cosmic strings $G \mu \simeq 10^{-6}$,
and $50 \lsim \Gamma \lsim 100$ 
(we have chosen
$\Gamma = 50$).  In Minkowski space, we took $\bar{\xi}=1$ and
$\zeta=5.0 \times 10^{-5}$
consistenly with the assumption of the time-independence of these
variables in a non-expanding universe.

As mentioned above and discussed in \dc{dp}, $0 \le K \le
0.1$ and we have taken
\beq
	K = 0.05.
\eeq
We will also consider the effect of varying $K$ about this value,
since it was shown in analytical analysis of \dc{dp} that the precise 
value of $K$ may influence whether or not a scaling solution can exist in 
the matter era.


Finally, observe that the probability $B(z',z-z';z)$ is very strongly
peaked about $z' = z-z' \simeq \zeta_c$.   In order to deal with such
rapidly changing functions in the region of $z=\zeta_c$ we found that
greatest numerical accuracy was achieved by choosing
\beq
	c_{max} = 10^{-4}.
\eeq
We then divided the $0 \leq y \leq 10^{-4}$ interval into $N$ equally
spaced intervals
(with $N=200$ or $N=500$), and used open integration routines and
adaptive mesh techniques in the $u$ direction to solve equation
(\dr{h4}).

We end this section with the following important observation.  By
choosing $c_{max}=10^{-4}$,  our numerical work can 
deal very accurately with {\em small} loops, but {\em cannot} deal
accurately with large loops.  The reason is the following.  Let us
assume that we work with $N=500$ intervals in the $y$ direction.
Then the last interval corresponds to $9.998 \times 10^{-5} \leq y
\leq 10^{-4}$, or alternatively using equation (\dr{zofy}) to 
 $ 0.0499 \leq z  \leq \infty$.  In other words, the last interval in $y$
contains all the loops with length $z \gsim \xi_c/50$, and the
evolution of these loops in time can clearly 
not be followed accurately using only one interval.  One
might then be concerned that inaccuracies with loops of length $z\gsim
\xi_c/50$ might lead to inaccuracies in the evolution of those with
smaller lengths since the rate equation (\dr{geqn}) is non-linear as
well as non-local.  This indeed will be true though it is difficult to 
estimate the size of these effects.  However, 
an expansion to lowest 
order in $z/\zeta_c$  of the rate equation (\dr{geqn}) shows 
that the small $z$ behaviour decouples from the large $z$ one 
\dc{dthesis} (though this is no longer true to higher order
in $z/\zeta_c$). 

We therefore conclude this section by saying that our numerics will be 
probing the small $z$ behaviour of the loop distribution $g(z,t)$ and
not the large $z \gg \zeta_c,\xi_c$ one which was the subject of
\dc{dp}.  This numerical study is therefore 
complementary to the analytic one of \dc{dp} 
rather than being at odds with it.

\section{Loops in Minkowski space time}\label{mink}

Given these comments, we begin by considering
the rate equation (\dr{ratemink}) in a non-expanding universe. (As
commented above for the
purpose of the numerics we work with 
(\dr{h4}) where only the four integrals on the RHS are kept and where
$u=t$).  

\subsection{Some analytical results}

In
Minkowski space we know the equilibrium solution of (\dr{ratemink}):
combining the principle of detailed balance with the form of the $A$ and $B$
coefficients gives  the
equilibrium (and hence stable) distribution as \dc{dp}
\beq
	n(\ell) = \frac{\tilde{\chi}}{\chi \bar{\xi}^{3/2}} 
\frac{e^{-\beta l}}{(l+\zeta)^{5/2}} =: c_{eq}\frac{e^{-\beta
l}}{(l+\zeta)^{5/2}} .
\label{eq:nesolution}
\eeq
Substitution of (\dr{eq:nesolution}) into (\dr{ratemink}) shows that
this is indeed a solution with $\beta$ arbitrary.  However, the
energy density in loops must be finite so that $\beta \geq 0$.

The solution (\dr{eq:nesolution}) is consistent with 
the statistical mechanics prediction
that in equilibrium \dc{stat}
\beq
	n(\ell)=ce^{-\beta \ell}\ell^{-5/2}
\label{eq:statmech}
\eeq
for $\ell \gg \zeta$.  It has also been shown \dc{stat} that at
$\beta = 0$ when the loops have a scale invariant distrubition, the
energy density in loops is equal to a critical value and a phase
transition occurs.   This is the Hagedorn transition for which the
loops network is unstable against the formation of an infinite string.

In \dc{dp}, the approach to the equilibrium solution
(\dr{eq:nesolution}) was studied using the rate equation
(\dr{ratemink}).  This was
done by substituting distributions of the form
\beq
	n(\ell,t)=\frac{e^{-\beta \ell}}{(\ell +\zeta)^{5/2}}
\sum\limits_{n=0}^{\infty} \frac{c_n(t)}{(\ell +\zeta)^n} 
\label{eq:initcond}
\eeq
with $c_n$ and $\beta$ are arbitrary parameters, into the rate
equation.  Equations for
$\dot{c}_n$ and $\dot{\beta}$ were then obtained by studying the $\ell \gg
\zeta$ limit of the rate equation.  Analysis of these equations
showed that a for an initial distribution of the form
(\dr{eq:initcond}), a {\em stable} equilibrium solution is reached
which is characterised by
\be
	c_0 \rightarrow c_{eq}, \qquad c_{n \geq 1} \rightarrow 0.
\eeq
Furthermore, since energy is conserved by the rate equation, 
$\beta$ was shown to change as
well, with its equilibrium value such that the energy
density at equilibrium is the same as the initial energy density.
This evolution was shown in figure 1 of \dc{dp}.

If the initial energy density in loops is larger than the critical
energy density, then one would expect from the results of statisical
mechanics that infinite strings are formed.  A plausibilty analysis
for this (based on the appearance of non-analytic terms when $\beta
\rightarrow 0$) was given in \dc{dp}.  In the present paper we are
concerned only with distributions that
have a total energy density less than that for the Hagedorn
transition.  Analysis of the case with energy greater than the critical 
one will be given in a future publication \dc{future} (see also
\dc{hthesis}).

\begin{figure}
\centerline{\psfig{file=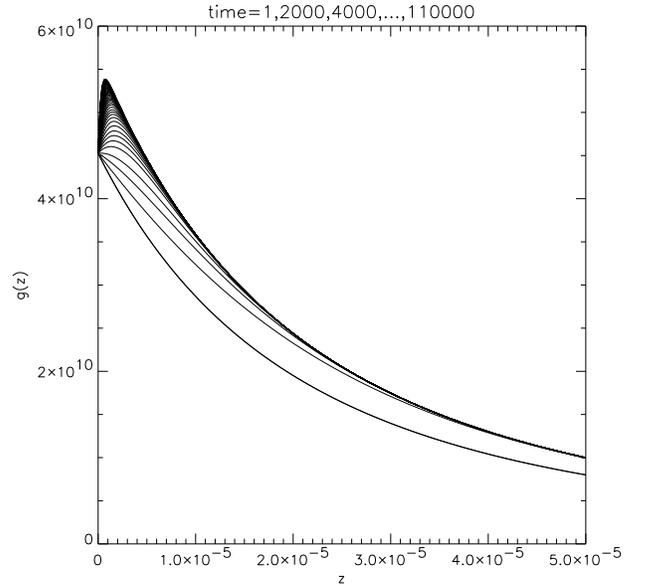,width=8 cm,angle=0}}
\caption{Initial conditions with $c_0=0.8<c_{eq}$, $c_{n>0}=0$,
$\beta_0=0.01$.  We have plotted 
the function $n(\ell,t)$ for times from $t=1$ to $t=110000$.}
\label{fig:ne1}
\end{figure}

\begin{figure}
\centerline{\psfig{file=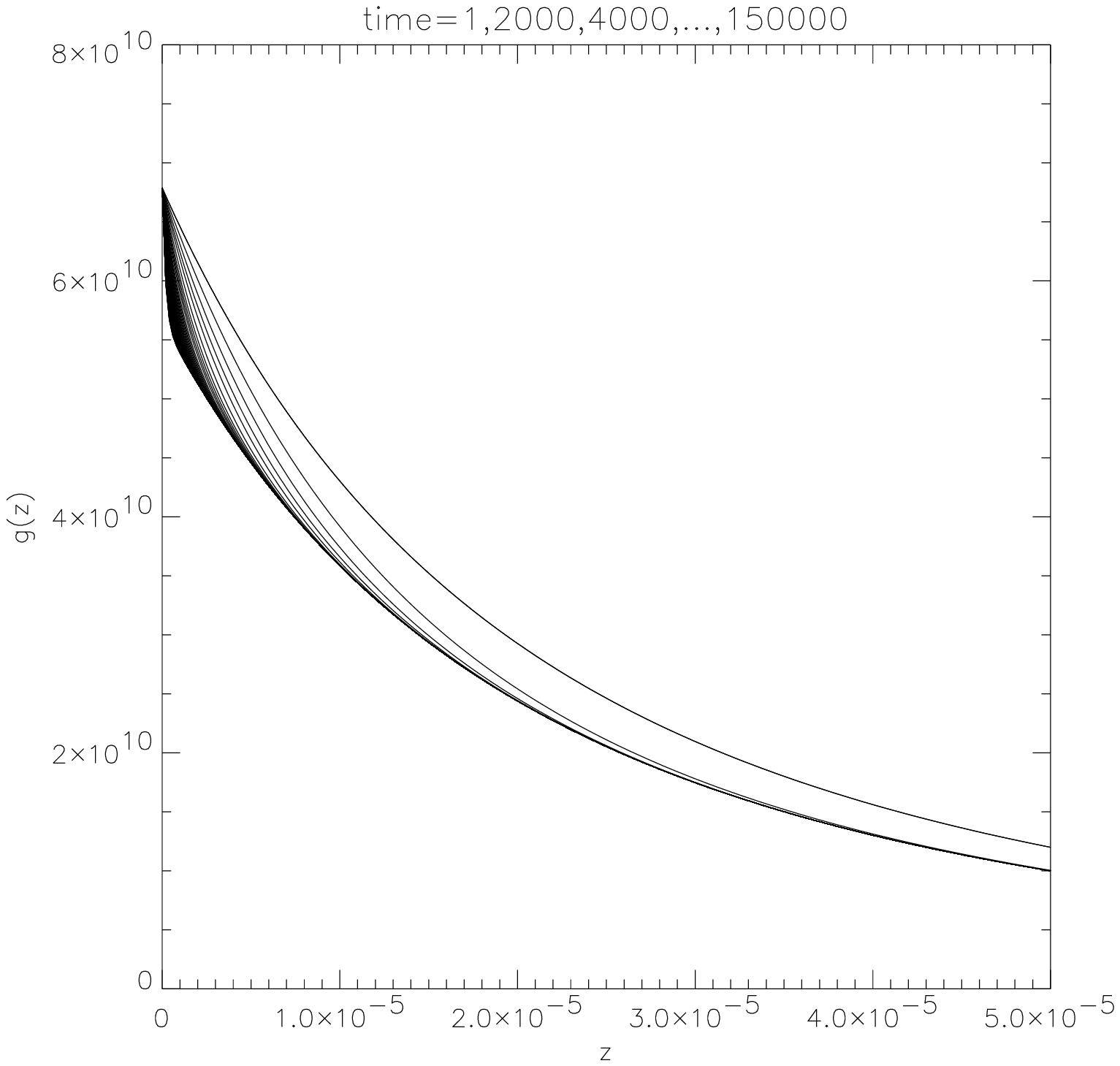,width=8 cm,angle=0}}
\caption{Initial conditions with $c_0=1.2>c_{eq}$, $c_{n>0}=0$,
$\beta=0.01$.  
We have plotted the function $n(\ell,t)$ for times from $t=1$ to
$t=110000$.} 
\label{fig:ne2}
\end{figure}

\begin{figure}
\centerline{\psfig{file=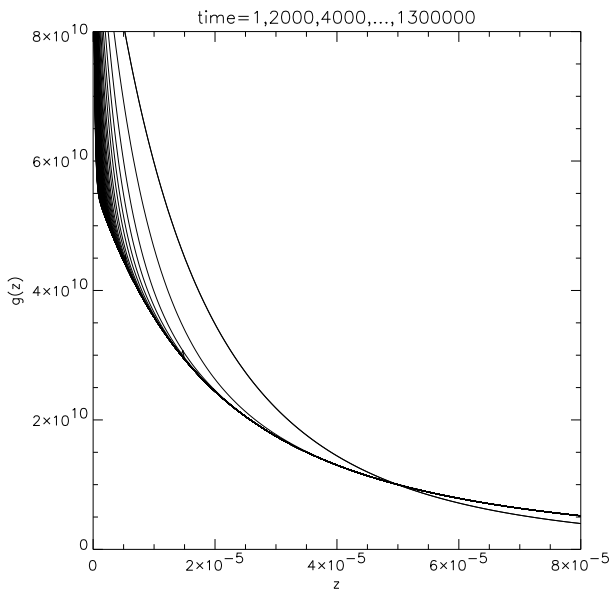,width=8 cm,angle=0}}
\caption{Initial conditions with $c_0=0$, $c_1 = 1.2>c_{eq}$, $c_{n>1}=0$,
$\beta=0.01$. 
We have plotted the function $n(\ell,t)$ for times $t=1$ to
$t=130000$.}
\label{fig:ne3}
\end{figure}

\subsection{Numerical results}

We have solved (\dr{ratemink}) numerically, starting 
with a variety of initial conditions  parametrised as 
in (\dr{eq:initcond}), but not necessarily close to
the equilibrium configuration, though always with subcritical energy density.

Figures (\ref{fig:ne1}) and (\ref{fig:ne2}) shows the time evolution
given the initial conditions, $c_0=0.8$ and $c_0=1.2$, respectively,
with $\beta=0.01$ and $c_{n > 0}=0$ in both cases.  Recall that with
our choice of parameters $c_{eq}=1$. 
In the first case
$c_0$ approaches $ c_{eq}$ from below, and in the second one $c_{eq}$ is 
approached from above, whilst in both cases the power dominant power
law behaviour 
is indeed $(\ell + \zeta)^{-5/2}$ as expected from (\dr{eq:nesolution}).
(In other words, all the $c_{n>0}$ remain zero for all times.)
It is difficult to see from these plots,
however, if $\beta$ changes as predicted since for these small values
of $\ell$, $e^{- \beta \ell} =1$ to very high accuracy (see further
comments on this below).


The only point of the distribution
which does reach the scaling solution is $n(0,t)$, which in fact does
not change. It can be shown analytically from (\dr{ratemink}) that 
the time derivative of $n$ at $\ell=0$ is indeed zero. There is very
little meaning in loops of size zero, so we do not comment further
on this exception to the equilibrium distribution.

In figure (\ref{fig:ne3}) the initial conditions have
 $c_0 = 0$, $c_1=1.2$, $c_{n>1}=0$ and $\beta=0.01$.  That is, this initial
distribution has a power law behaviour of $(\ell + \zeta)^{-7/2}$.
Despite that, the plot shows that as the network evolves, 
$c_1 \rightarrow 0$ and $c_0 \rightarrow 1$ so that the system evolves 
to its equilibrium distribution (though again it is difficult to
comment on $\beta$).

The stability of this solution with $c_0=c_{eq}$ and $c_{n>0}=0$ 
is very important.  Previous
proofs of stability have been confined to \emph{linear}
perturbations.  No analytical evidence has been given for stability in 
the nonlinear regime.  Our results bear strong evidence that the
solution is reached with \emph{any} initial conditions, and thus that
it is stable also in the non-perturbative regime.  

An important test for the code is to check whether the total loop
length density (energy) stays constant.  The simulations show that the more
sampling points (the larger $N$), 
the better the conservation law is satisfied.  We  
decided to accept sampling densities leading to  
mass variations of $\sim 5 \% $.

\section{Expanding Universes}\label{expand}

\subsection{Analytical results}

In \dc{dp} the asymptotic behaviour
of solutions to the rate equation (\dr{geqn}) were searched for. More
specifically, using the ansatz  
\beq
g(z,t) = \theta(z) \frac{e^{-\beta(t) z}}{(z+\zeta)^{\frac{5}{2}}}
\sum\limits_{n=0}^{\infty} \frac{c_n(t)}{(z+\zeta)^n}
\label{eq:nansatz},
\eeq	
substituting into the rate equation (\dr{geqn}), and assuming $(z \gg
\bar{\xi}_c \gg \zeta_c$) led to equations for $\dot{\beta}$ and
$\dot{c_n}$.  Analysis of these led to the conclusion that, at least
to order $n=1$, a scaling solution exists in the radiation era but not 
in the matter era.   Furthermore, the radiation era solution was
thought to be unstable.  That is, depending on the initial energy
density in the loops, these were either thought to disappear or
infinite strings were thought to be formed.  In the matter era no
stable solution was found and for much of parameter space the loops
were expected to disappear \dc{dp}.


Before presenting our numerical analysis of this expanding universe case, 
we introduce the concept of the mean loop length since
this will be useful for the analysis of our numerical solutions.
Recall that by definition, scaling of the energy density of the loop
network means that 
\begin{equation}
	\frac{dE_{sc}}{dt}=0,
\label{eq:scalingcriteria}
\end{equation}
where
\be 
E_{sc} = \mu \int_0^{\infty} z g(z,t) dz.
\dle{Escdef}
\ee
As was shown in \dc{dthesis}, $E_{sc}$ satisfies
\beq
	t\frac{dE_{sc}}{dt}=(2 + K - 3/p)E_{sc}(t) - \Gamma G \mu N_{sc}(t)
\label{eq:debydt}	
\eeq
where $N_{sc}(t)$ is the total number of loops in scaling variables,
\beq
	N_{sc}(t)=\int_0^{\infty} g(z,t)dz.
\eeq
As commented in \dc{dp}, we see that the assumption of scaling lengths 
$\bar{\xi}$ and $\zeta$ does not automatically imply that the energy
density in loops also scales in an expanding universe:  this will only 
happen if $N_{sc}$ and $E_{sc}$ reach values such that the RHS of
(\dr{eq:debydt}) vanishes.  If that is the case, then the 
mean length $\lav z\rav$ is given by
\beq
	\lav z \rav = \frac{\int_0^{\infty}zg(z,t)dz}{\int_0^{\infty}g(z,t)dz}
= \frac{E_{sc}(t)}{N_{sc}(t)} = \frac{\Gamma G \mu}{2 + K - 3/p}.
\label{eq:scalingmean}
\eeq	
With our choice of
parameters 
\ba
	\lav z \rav_{rad} &\sim  & 1 \times 10^{-4} = 2 \zeta_c
\nonumber
\\
\lav z\rav_{matt}
&\sim & 1 \times 10^{-3} = 20 \zeta_c 
\label{eq:meanz}
\ea

\subsection{Numerical Results}

\subsubsection{Radiation era}

\begin{figure}
\centerline{\psfig{file=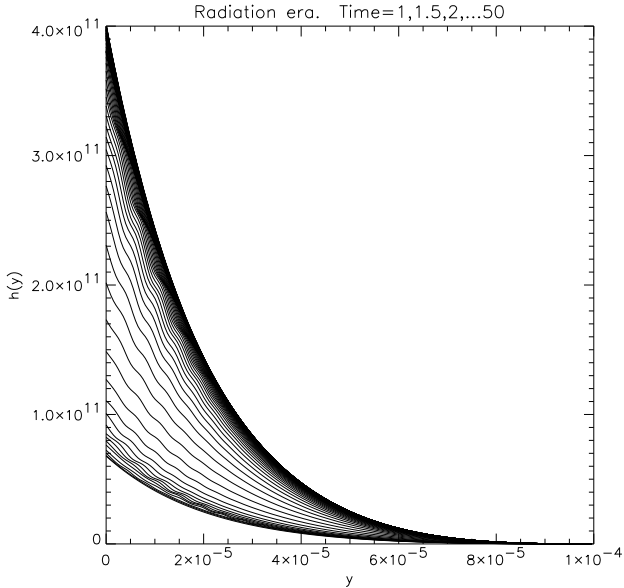,width=8 cm,angle=0}}
\caption{Evolution of the network in the radiation era.  Initial
conditions were $c_0 = 1.2$, $c_{n>0}=0$ and $\beta = 10^{-2}$.
We see that the 
distribution after 10 to 20 logarithmic times reaches a stable scaling
solution.  We have plotted the function $h(y)$.}
\label{fig:rad1}
\end{figure}
\begin{figure}
\centerline{\psfig{file=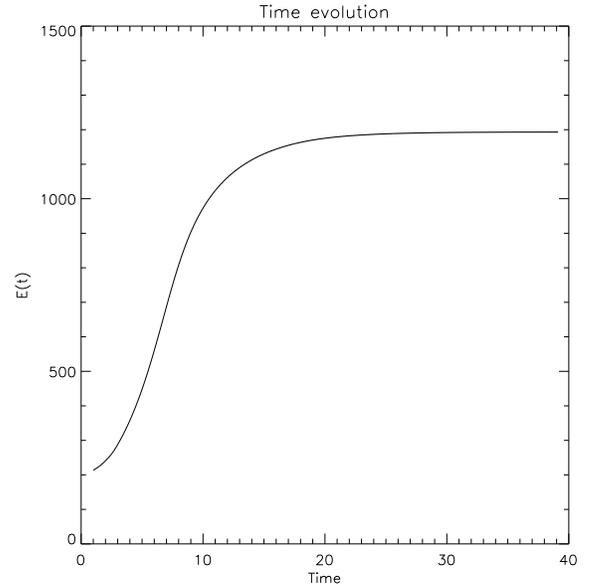,width=8 cm,angle=0}}
\caption{Plot of the scaling energy density, $E(u)$ during the evolution of the
network in the radiation era.  Initial conditions were $c_0 = 1.2$,
$\beta = 10^{-2}$ and $c_{n>0}=0$.  The energy density 
increases rapidly just after start, and firmly, after about 20
logarithmic times, approaches the scaling value from below.}
\label{fig:radenergy}
\end{figure}
We gave the network initial conditions of the form (\ref{eq:nansatz}). 
but with various values for  $c_n$ and $\beta$.
In figure (\ref{fig:rad1}) we show the time evolution of $h(y,u)$ given the
initial conditions $c_0=1.2$, $c_{n>0}=0$ and $\beta = 10^{-2}$.
The time
range is $u =
1,1.5,..,50$.  Figure (\ref{fig:radenergy}) shows the scaling 
energy density (defined in (\dr{Escdef}))
as a function of time.  We see that the energy density after 10
- 20 logarithmic times approaches a stable scaling value from
below.

\begin{figure}
\centerline{\psfig{file=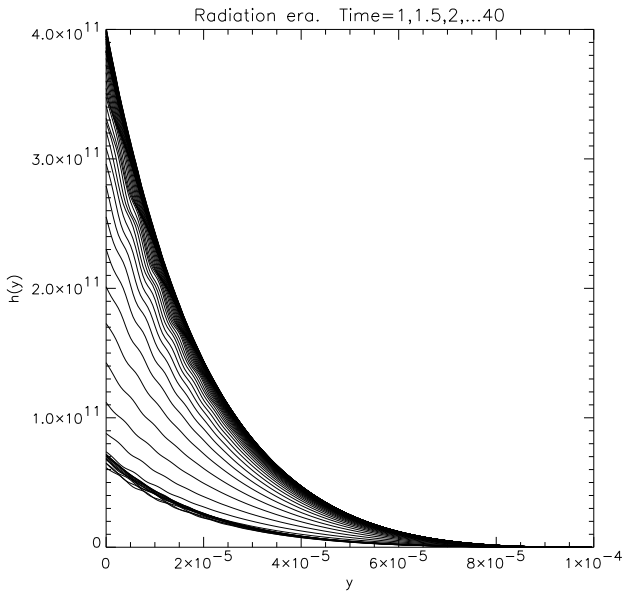,width=8 cm,angle=0}}
\caption{Evolution of the network in the radiation era.  Initial
conditions were $c_0 = 1.2$, $\beta = 10^{2}$ and $c_{n>0}=0$.
We see that the distribution after 12 to 15 logarithmic times reaches
a stable scaling solution.  We have plotted the function $h(y)$.}
\label{fig:rad2}
\end{figure}

\begin{figure}
\centerline{\psfig{file=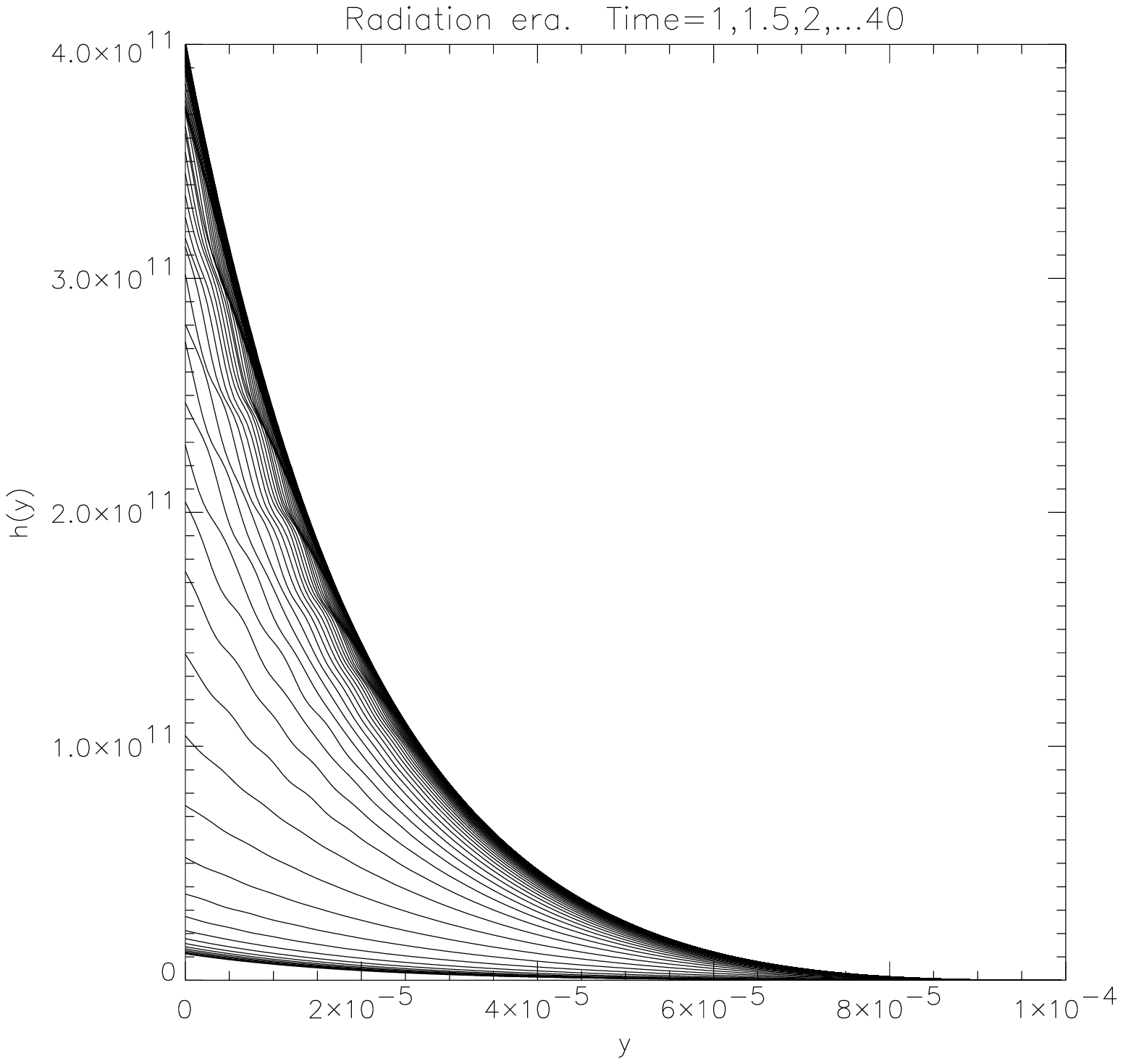,width=8 cm,angle=0}}
\caption{Evolution of the network in the radiation era.  Initial
conditions were $c_0 = 0.2$, $\beta = 10^{-2}$ and $c_{n>0}=0$.
We see that the distribution after about 20 logarithmic times reaches
a stable scaling solution.  We have plotted the function $h(y)$.}
\label{fig:rad3}
\end{figure}

\begin{figure}
\centerline{\psfig{file=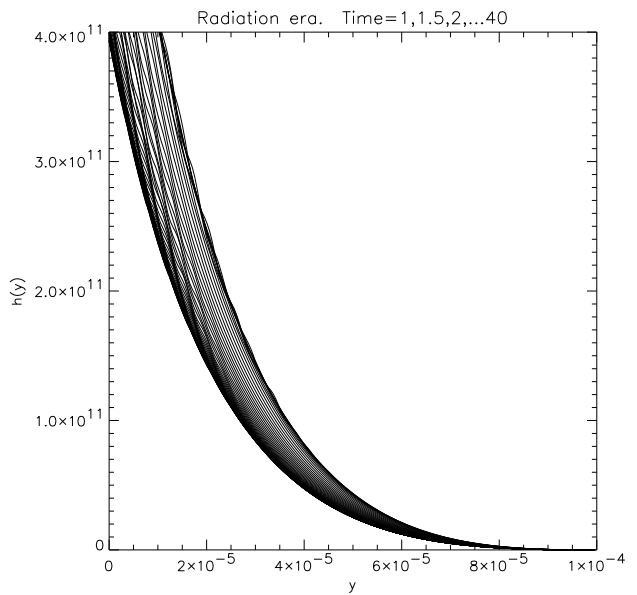,width=8 cm,angle=0}}
\caption{Evolution of the network in the radiation era. Initial
conditions were $c_0 = 12$, $\beta_0 = 10^{-2}$ and $c_{n>0}=0$,
which means we start out well above the scaling solution.  We see that
the distribution after about 20 logarithmic times reaches a stable scaling
solution.  We have plotted the function $h(y)$.}
\label{fig:rad4}
\end{figure}

In figures (\ref{fig:rad2}) and (\ref{fig:rad3}) we show the time
evolution given two different initial distributions both with
energies below the scaling value. 
%
It appears that these reach the same scaling distribution as that of
figures (\ref{fig:rad2}) and (\ref{fig:rad3}) --- we shall comment on
the form of this distribution below.
Figure (\ref{fig:rad4}) shows time
evolution starting out with an initial distribution with energy density higher
than the scaling value, $c_0=12, \beta=0.01$ and $c_{n>0}=0$.  Again the
same scaling solution seem to be approached from above.
The same solution is also found if we set $c_0=0$ but other of the
$c_{n>0} \neq 0$.  These results provide evidence
that the small loops reach a unique stable scaling solution 
in the radiation epoch,
and that this solution is an attractor within a very large basin
of initial conditions. 

\begin{figure}
\centerline{\psfig{file=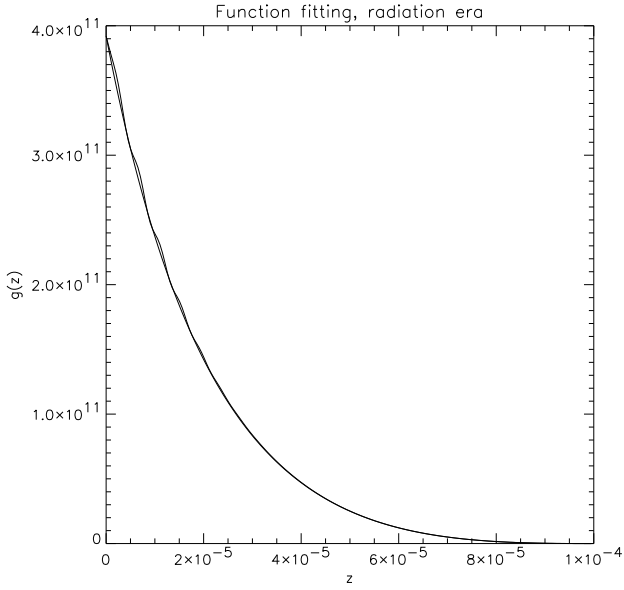,width=8 cm,angle=0}}
\caption{Attempt to fit the radiation scaling solution to the ansatz,
$g(z,t)=ce^{-\beta z}/(z+\zeta)^{5/2}$.  As we can see this is a nearly
perfect fit.  The constants found are, $c= 6.9$ and $\beta = 34$} 
\label{fig:radfit}
\end{figure}

We now attempt to characterise more thoroughly the scaling 
solution. Its  energy density is given by  
\beq
	E_{sc} \approx 1.2\times 10^3
\eeq
We could try to fit its shape with the ansatz 
$g(z)=ce^{-\beta z}/(z+\zeta)^{5/2}$. This is the leading order term
in the ansatz (\dr{eq:nansatz}) 
which was used in \dc{dp} for looking at large loops.
We used a  least squares method and the result 
is plotted in Figure (~\ref{fig:radfit}): clearly
this function is a good approximation in the 
radiation epoch.
The optimal combination of $c$ and $\beta$ is
\beq
	c_{rad} \approx 6.9, \qquad \beta_{rad} \approx 34
\eeq
(Though the work of \dc{dp} was probing a very different regime, 
there the (un-stable) scaling solution had values $c_{rad} = 7.1$ and
$\beta_{rad} = 108$.)  

Finally,
the scaling solution has a mean loop length
\beq
	\lav z \rav_{rad,numerical} \approx 1.6 \times 10^{-4}
\eeq
in good agreement  with the theoretically predicted value of 
$\lav z
\rav \approx 1 \times 10^{-4}$ from equation (\dr{eq:meanz}).

\begin{figure}
\centerline{\psfig{file=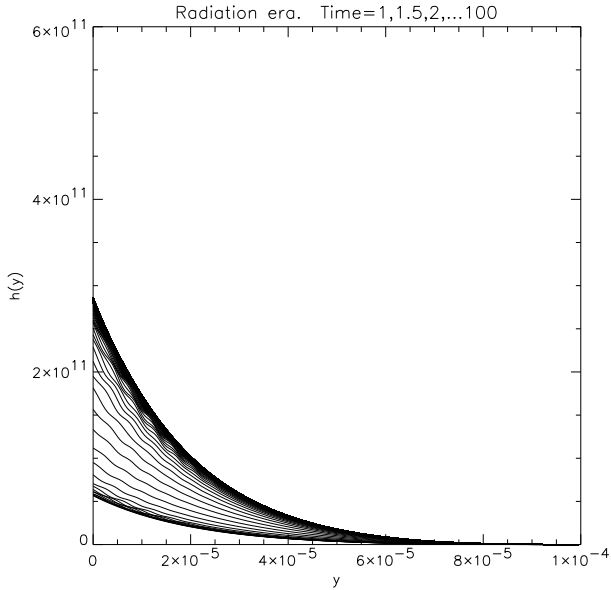,width=8 cm,angle=0}}
\caption{The scaling solution in the radiation era, given a new value
for $K = 0.025$.  We see that the solution is significantly smaller
than the previous one, with $E_{rad} \approx 0.9 \cdot 10^3 $ and
$\lav z \rav = 1.70\cdot 10^{-4}$} 
\label{fig:newk}
\end{figure}

In an attempt to understand how the solution depends on the parameters
involved, we also show results corresponding to a different
value for $K$.  Figure (~\ref{fig:newk}) shows how a scaling solution
is still approached, but now with a lower energy density, $E_{sc} \sim
9 \times 10^2$.  This is understandable, since a smaller $K$ could represent
either a larger loss from gravitational radiation in the large loop
limit or a greater loss from redshift of velocities.

\subsubsection{Matter epoch}

We performed a similar analysis in the matter epoch. 
The analytical analysis of \cite{dp} predicts that no scaling solution
exists for large $z$ in this case.  Whilst this may well be the case, our 
numerical results show that a scaling solution seems to be reached for 
small $z$.

\begin{figure}
\centerline{\psfig{file=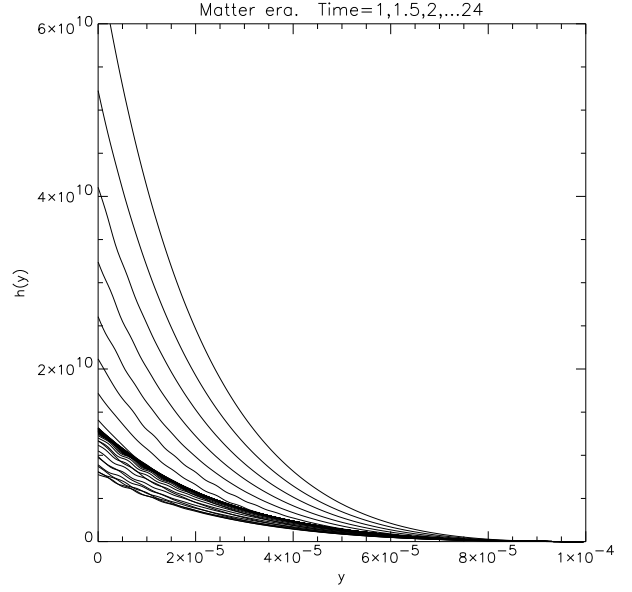,width=8 cm,angle=0}}
\caption{Evolution of the network in the matter era.  Initial
conditions were $c_0 = 1.2$, $c_{n>0}=0$ and $\beta = 10^{-2}$.  We see that the 
distribution immediately after the start starts decreasing and
reshaping.  The distribution goes a bit below the scaling solution,
but catches up.  The loop distribution stabilises after about 10-15
logarithmic times, but at value far lower than the radiation era.  We
have plotted the function $h(y)$.} 
\label{fig:matt1}
\end{figure}
\begin{figure}
\centerline{\psfig{file=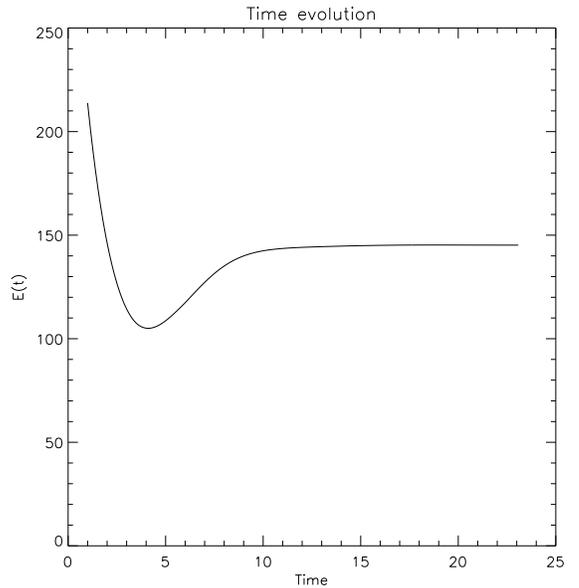,width=8 cm,angle=0}}
\caption{Plot of the scaling energy density during the evolution of the
network in the matter era.  Initial 
conditions were $c_0 = 1.2$, $c_{n>0}=0$ and $\beta = 10^{-2}$.  The energy density
decreases rapidly just after start, but then approaches the scaling
value from below.} 
\label{fig:mattenergy}
\end{figure}
Initial conditions of the form (\ref{eq:nansatz}) were again used. 
Figure (~\ref{fig:matt1}) shows the time evolution of $h(y,u)$ in the
matter era given
initial conditions, $c_0=1.2$, $\beta=0.01$ and $c_{n>0}=0$.  The
corresponding scaling energy density is plotted as a function of
logarithmic time, $E(u)$ in 
figure (~\ref{fig:mattenergy}). It shows how an energy 
density over the scaling value decreases rapidly, overshoots,
and then increases to asymptote towards the scaling value after
about 10 expansion times. 

\begin{figure}
\centerline{\psfig{file=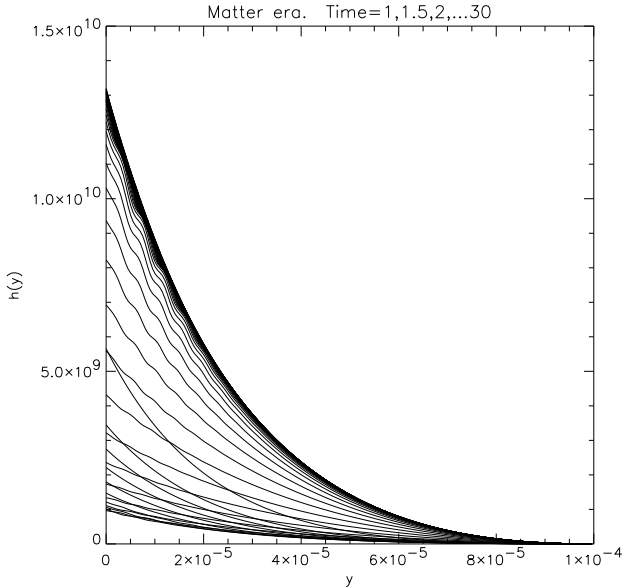,width=8 cm,angle=0}}
\caption{Evolution of the network in the matter era.  Initial
conditions were $c_0 = 0.1$, $c_{n>0}=0$ and $\beta = 10^{-2}$.  We see that the 
distribution immediately after the start starts decreasing and
reshaping.  The distribution then increases until it reaches the
scaling distribution for the matter era.  The loop distribution
stabilises after about 10-15 logarithmic times.  We
have again plotted the function $h(y)$.} 
\label{fig:matt2}
\end{figure}
Figure (~\ref{fig:matt2})
shows the time evolution of $h(y,u)$ 
given an initial distribution with an energy density lower than the
scaling value.  The function immediately starts reshaping, and after
some time the same scaling distribution is found.
We considered also initial configurations  with different values 
$c_{n>0} \neq 0$, and found that they approach 
the same scaling solution.  As in the radiation era the same
scaling solution is always reached, starting from a large sample
of possible initial conditions.  Again though we are not very
sensitive to different values of $\beta$.

The stable scaling solution in the matter epoch is rather different
from the radiation epoch scaling solution. Its scaling energy density 
is nearly 10 times smaller 
\beq
	E \approx 1.5\times 10^2,
\eeq
Its shape is also very different. The mean scaling loop length $\lav z\rav$,
is larger in the matter epoch, as predicted. We found 
\beq
	\lav z \rav_{matt,numerical} \approx 0.5 \times 10^{-3}.
\eeq
The difference between $\lav z \rav_{rad}$ and $\lav z\rav_{matt}$ is not
as large as theoretically predicted, but shows the right tendency.

\begin{figure}
\centerline{\psfig{file=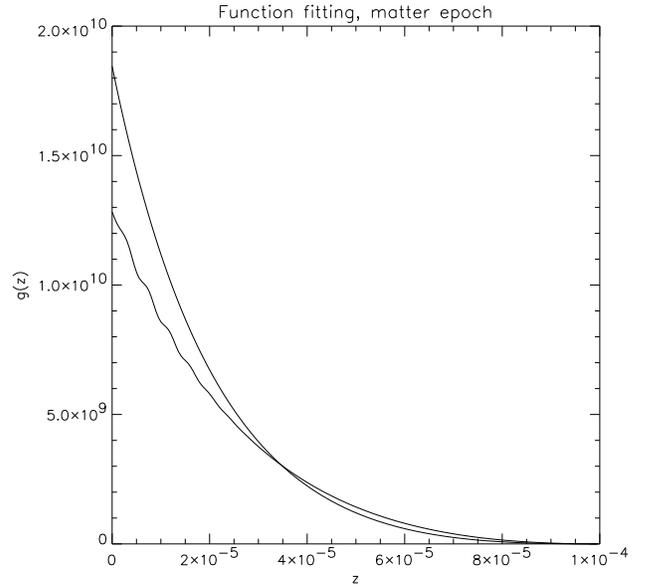,width=8 cm,angle=0}}
\caption{Attempt to fit the matter scaling solution to the ansatz,
$g(z,t)=ce^{-\beta z}/(z+\zeta)^{5/2}$.  As we can see the functions
are not compatible.} 
\label{fig:functionfit2}
\end{figure}

Finally we tried to fit the scaling solution to an ansatz 
$g(z)=ce^{-\beta z}/(z+\zeta)^{5/2}$. 
Again we have used a least squares method.
The result is plotted in (~\ref{fig:functionfit2}). Clearly
our scaling solution does not fit this ansatz.  However, it is not
difficult to see that a distribution of the form $g(z,t)=ce^{-\beta
z}/(z+\zeta)^{n}$ with $n<5/2$ would fit our scaling distribution better.

\section{Conclusions}

\label{conc}

We re-examined the evolution of networks of
cosmic strings consisting solely of loops. Our aim was to provide
full numerical solutions to the rate equation first given in
\cite{dp}, though in fact we were only able to probe accurately small loops.

In Minkowski space time we found that for energies lower than the critical
energy the system evolves towards a scaling solution. This happens for
a large variety of initial conditions not necessarily in the vicinity
of the scaling solution. The results obtained with our numerical code
agree with the statistical mechanics prediction, as well as with 
the analytical predictions made in \cite{dp}.  However we also 
predict that the scaling solution is stable against non linear 
perturbations, as long as the critical mass is never exceeded.
For total energies larger than the critical we were unable to obtain
reliable results, though infinite strings are expected to form.

In expanding universes and for small loops, 
a scaling solution was found both in the
radiation and matter dominated epochs.  The shape of the
solution is quite different in the two eras.  The gradients of the
distributions for small $z$ agree with an analytical analysis of the
rate equation for small $z$ carried out in
\dc{dthesis}.  The scaling energy densities
are $E_{rad} \sim 1200$ and $E_{matter} \sim 150$, whereas the average
scaling loop lengths, $z=\ell/t$, are $\lav z_{rad}\rav \approx
10^{-4}$, and $\lav z_{matt} \rav \approx 0.5 \times 10^{-3}$.

Our results complement the findings in \cite{dp} 
for radiation and matter dominated Universes. 
In both these cases we found evidence for scaling solutions containing 
many small loops.  Thus there will always be a scaling solution in
these eras, though from the results of \dc{dp} these may or may not
contain many loops with length larger than the horizon.


We close with a description of three promising research avenues
which our work may have prompted. Firstly
our results are very encouraging for a loop dominated
scenario of structure formation. Our discovery of  scaling solutions
for small loops
ensures that no cosmological cataclysm happens in these scenarios.
Undoubtedly the unequal time correlators, required for computing 
structure formation, are very different in loop networks and in 
the usual networks
dominated by long strings. The predicted CMB anisotropy and
galaxy  power spectra will therefore be very different. One wonders
if a better fit to the data might be achieved in these models.

Before this idea can be carried out, we should investigate 
the radiation to matter transition which 
 is a crucial stage in structure formation.  We believe
that a long lived transient is likely to set in. 

It would also be interesting to reexamine our results 
with a dynamical equation allowing for the coexistence of loops
and infinite strings. 
We reserved to a future
publication the presentation of numerical solutions ~\cite{future}. 

{\bf Acknowledgements}  
We are very grateful to Tom Kibble for invaluable help. 
We acknowledge  financial support from the Royal Society 
(JM), PPARC (DS).

\end{document}